\documentclass[12pt]{article}
\usepackage{amsfonts,amssymb,epsfig,amsmath}
\usepackage{color}
\renewcommand{\text}[1]{#1}

\newcommand{\be}{\begin{equation}}
\newcommand{\ee}{\end{equation}}
\newcommand{\ben}{\begin{displaymath}}
\newcommand{\een}{\end{displaymath}}
\newcommand{\bea}{\begin{eqnarray}}
\newcommand{\eea}{\end{eqnarray}}
\newcommand{\bean}{\begin{eqnarray*}}
\newcommand{\eean}{\end{eqnarray*}}
\newcommand{\nn}{\nonumber \\}
\newcommand{\ba}{\begin{array}}
\newcommand{\ea}{\end{array}}
\newcommand{\bi}{\begin{itemize}}
\newcommand{\ei}{\end{itemize}}

\renewcommand{\theequation}{\arabic{section}.\arabic{equation}}
\def\theequation{\thesection.\arabic{equation}}

\def\l{\lambda}

\def\g{\gamma}

\def\g{\gamma}
\def\e{\epsilon}

\def\e{\epsilon}

\begin{document}

\makeatletter
\renewcommand{\theequation}{\thesection.\arabic{equation}}
\@addtoreset{equation}{section}
\makeatother

\begin{titlepage}

\vfill
\begin{flushright}
KIAS-P10035
\end{flushright}

\vfill

\begin{center}
   \baselineskip=16pt
   {\Large\bf  On the generality of the LLM geometries in M-theory }
   \vskip 2cm
     Eoin \'O Colg\'ain$^{1}$, Jun-Bao Wu$^{1,3}$ \& Hossein Yavartanoo$^{2}$
       \vskip .6cm
             \begin{small}
      \textit{Korea Institute for Advanced Study$^{1}$, \\
        Seoul 130-722, Korea \\ \vspace{2mm}
        Department of Physics, Kyung Hee University$^{2}$, \\Seoul 130-701, Korea \\ \vspace{2mm} Institute of High Energy Physics$^{3}$, \\
Chinese Academy of Sciences, \\
100049 Beijing, P. R. China}
        \end{small}\\*[.6cm]
\end{center}

\vfill
\begin{center}
\textbf{Abstract}\end{center}

\begin{quote}
In this note we revisit the Lin, Lunin, Maldacena (LLM) class of $d=11$ supergravity solutions with symmetry $SO(6) \times SO(3) \times R$, but generalise to allow for all fluxes consistent with the isometries. Using the Killing spinor equation, we prove there are no supersymmetric geometries with additional fluxes beyond the LLM ansatz.  In addition, the LLM relationship between Killing spinors, $\e_- = - \g_5 \e_+$, may be seen as a consequence of identifying two Killing directions identified through the Killing spinor equation corresponding to candidate $R$-symmetry directions.
\end{quote}
\vfill

\end{titlepage}

\section{Introduction}
 A reputed most general gravity solution of eleven dimensional supergravity which preserves $\mathcal{N} = 2$ superconformal symmetry was constructed in \cite{LLM}. Despite the commonly held belief that it is the most general class of solutions, a question exists over the existence of solutions with an additional flux term. If such a supersymmetry preserving flux term could be incorporated into the ansatz, then it raises pressing questions about the field theory interpretation, and with the recent interest in dual geometries (e.g. \cite{GM}), a thorough analysis is well overdue. In this note we address that question by adopting the same $S^5 \times S^2 \times \mathcal{M}_4$ ansatz  as LLM\footnote{The $S^5$ is easily analytically continued to $AdS_5$ as explained in \cite{LLM}.}, but in contrast we treat the two Killing spinors $\e_+, \e_-$  on the spacetime $\mathcal{M}_4$ independently to see how the Killing spinor equation (KSE) constrains the geometry. We begin by reviewing the arguments and assumptions of LLM \cite{LLM}.

We recall that \cite{LLM} argued that the presence of a four-form flux term $\mathcal{G}$, proportional to the volume of the external spacetime $\mathcal{M}_4$, should be set to zero to avoid spheres shrinking in an irregular fashion. The reason being that flux density would diverge at the points where the spheres shrink. Once this simplification is adopted, it was subsequently noted that the Killing spinor equations simplify into two decoupled systems for the linear combinations $\e_- \pm \g_5 \e_+$. A choice of sign $\e_- = - \g_5 \e_+$ then leaves one with a single independent spinor and generates the geometry through the powerful techniques of G-structures (see \cite{M6, M5} for similar examples with less supersymmetry and $d=4$ dual SCFTs). The authors finally show that starting from the LLM class of solutions, one cannot perturbatively turn on $\mathcal{G}$, however this does not rule out geometries where $\mathcal{G}$ exists and cannot be set to zero. The aim of this work is to challenge the generality of these assumptions by relaxing them.

Our work builds on the KSE decomposition presented in \cite{LLM} by retaining $\mathcal{G}$ and treating $\e_+$ and $\e_-$ initially in an independent fashion, though of course they will be related through supersymmetry. While our approach does have some overlap with the work of \cite{M5}, we follow largely in the footsteps of a classification of $AdS_3 \times S^2$ geometries with $SU(2)$-structure dual to chiral $\mathcal{N} = (4,0)$ SCFTs \cite{ads3M6}, though here the space to be determined is of dimension four and not six, leading to a simplification and more generality. In other words, as we are working in four-dimensions, one can assume that a generic Killing spinor $\e_+$ (non-chiral, non-Majorana) may be related to another $\e_-$ simply through complex functions $a_i$ such that
\be
\e_- = \sum_{i=1}^{4} a_i \eta_{i},~~\mbox{where}~~\eta_{i} \in \{ \e_+, \g_5 \e_+, \e_+^{c}, \g_5 \e_+^c\}.
\ee

Our strategy for determining the $a_i$ will hinge largely on the algebraic constraints arising from the KSE. These place strong constraints on the scalar and vector bilinears one may construct from the Killing spinors. Though one may derive many constraints, the ones we present later are sufficient for our purposes. In the present setting of the LLM ansatz one also benefits from a particular linear combination of the algebraic constraints in which a two-form flux term $\mathcal{F}$ on $\mathcal{M}_4$ does not appear \cite{LLM}. We will see in the first half of this note that if one insists on $\mathcal{G} \neq 0$, then there is no non-trivial solution to the KSE. However, when one sets $\mathcal{G}$ to zero, it is possible to ask if the choice $\e_- = - \g_5 \e_+$ is the most general consistent with the ansatz and supersymmetry. We show that if the two Killing directions we note later in the text are identified\footnote{See \cite{Donos} for a recent study of LLM geometries with an additional $U(1)$ isometry.},  so that they correspond  to the $R$-symmetry, as in the case of LLM, then this relationship is the only possibility.

\section{Warm-up}
We begin this section by reviewing the ansatz of LLM \cite{LLM}. The $d=11$ supergravity ansatz may be written as a warped product of $S^5$, $S^2$ and a non-compact $\mathcal{M}_4$ spacetime,
\bea
ds^{2} &=& e^{2 \lambda} \left[ \frac{1}{m^2} d \Omega_5^2 + e^{2 A} d \Omega_2^2 + ds^2_{\mathcal{M}_4} \right], \nn
F^{(4)} &=& \mathcal{G} + vol(S^2) \wedge \mathcal{F},
\eea
where $\lambda, A$ are functions only of the coordinates on $\mathcal{M}_4$ and $\mathcal{F}$ and $\mathcal{G}$ are respectively 2-forms and 4-forms on $\mathcal{M}_4$.  The Bianchi identity will be satisfied through the closure of these forms, while the flux equations of motion are
\bea
d (e^{3 \lambda + 2 A} *_4 \mathcal{G} ) &=& 0,  \nn
d (e^{3 \lambda -2 A} *_4 \mathcal{F} ) &=& 0.
\eea
Provided these equations of motion are satisfied, it is at this stage well established that  Einstein equations follow from integrability of the KSE \cite{int}, a fact that guarantees the existence of a genuine $d=11$ supergravity solution.

We next reproduce the results of the Killing spinor decomposition, the details of which may be found in appendix F of \cite{LLM},  to get the following equations in our notation:
\bea
\left(\g^\mu\partial_\mu\lambda-\frac{i}6e^{-6\lambda-2A} \mathcal{I}\g_5\right)\e_{\pm}
& \mp & \left(\frac1{12}e^{-3\lambda-2A}\g^{\mu\nu} \mathcal{F}_{\mu \nu}
+m\g_5 \right)\e_{\mp}=0,\label{const1}\\
\left(\pm ie^{-A}\g_5+\g^\mu\partial_\mu A\right)\e_{\pm} & \pm &
\left(m\g_5+\frac14 e^{-3\lambda-2A}\g^{\mu\nu} \mathcal{F}_{\mu\nu} \right)\e_{\mp}=0,\label{const2}\\
\left( \nabla_\mu-\frac{i}4\g_5\g_\mu e^{-6\lambda-2A} \mathcal{I}
\right)\e_\pm & \pm & \left( \frac{m}2\g_\mu\g_5+\frac14 e^{-3\lambda-2A} \g^\nu
\mathcal{F}_{\mu \nu}  \right)\e_{\mp}=0.\label{diff3} \eea A
linear combination of (\ref{const1}) and (\ref{const2}) leads to an algebraic condition independent of $\mathcal{F}$ \be
\left( \g^\mu\partial_\mu (3\lambda+A)\pm
ie^{-A}\g_5-\frac{i}2e^{-6\lambda-2A} \mathcal{I} \g_5 \right)\e_\pm \mp
2m\g_5\e_\mp=0. \label{const4} \ee

Note as in LLM, we have also imposed the equations of motion on $\mathcal{G}$ through the redefinition
\be
\mathcal{G} = e^{-3 \lambda -2 A} \mathcal{I} vol(\mathcal{M}_4),
\ee
where $\mathcal{I}$ is now simply the constant appearing in the above equations.

Before proceeding with the analysis, consistent with the conventions of LLM (appendix A)\footnote{We are grateful for correspondence with Oleg Lunin on this matter. }, we start by making an exhaustive list of the scalars
\bea S_1&=&\frac{i}2(\bar{\e}_+\e_++\bar{\e}_-\e_-), \quad S_2=\frac{i}2(\bar{\e}_+\e_+-\bar{\e}_-\e_-), \quad S_3=\bar{\e}_+\e_-, \nn T_1&=&\frac12(\bar\epsilon_+\g_5\epsilon_++\bar\epsilon_-\g_5\epsilon_-), \quad
T_2=\frac12(\bar\epsilon_+\g_5\epsilon_+-\bar\epsilon_-\g_5\epsilon_-), \quad
T_3=\bar\epsilon_+\g_5\epsilon_-, \nn
U_1 &=& \bar{\e}_{+}^{c} \e_{-}, \quad U_2 = \bar{\e}_{+}^{c} \g_5 \e_-,
\eea
and the vectors
\bea
K^{1}_{\mu} &=& \tfrac{1}{2} ( \bar{\e}_+ \g_{\mu} \e_+ + \bar{\e}_- \g_{\mu} \e_- ), \quad K^{2}_{\mu} = \tfrac{1}{2} ( \bar{\e}_+ \g_{\mu} \e_+ - \bar{\e}_- \g_{\mu} \e_- ) \nn K^{3}_{\mu} &=& \tfrac{1}{2} ( \bar{\e}_+ \g_5 \g_{\mu} \e_+ + \bar{\e}_- \g_5 \g_{\mu} \e_- ), \quad K^{4}_{\mu} = \tfrac{1}{2} ( \bar{\e}_+\g_5 \g_{\mu} \e_+ - \bar{\e}_- \g_5 \g_{\mu} \e_- ), \nn
K^{5}_{\mu} &=& \tfrac{1}{2} ( \bar{\e}_+^c \g_{\mu} \e_+ + \bar{\e}_-^c \g_{\mu} \e_- ), \quad K^{6}_{\mu} = \tfrac{1}{2} ( \bar{\e}_+^c \g_{\mu} \e_+ - \bar{\e}_-^c \g_{\mu} \e_- ), \nn
K^7_{\mu} &=& \bar{\e}_+ \g_{\mu} \e_-, \quad K^8_{\mu} = \bar{\e}_+ \g_5 \g_{\mu} \e_-, \nn
K^9_{\mu} &=& \bar{\e}_+^c \g_{\mu} \e_-, \quad K^{10}_{\mu} = \bar{\e}_+^c \g_5 \g_{\mu} \e_-.
\eea
that will appear later. The omitted scalar and vector bilinears may be shown to be zero by employing the symmetry properties of the gamma matrices. It should be borne in mind that depending on the choice of $\e_+, \e_-$, some of these bilinears will be trivially zero. As we are also working in four dimensions, we expect that these vectors are related - they cannot be independent. In fact, some relationships between them may be deduced by playing around with the algebraic constraints. The results of some of those efforts we will see later.

As the sight of all these bilinears may be a bit overwhelming, it is helpful to orient ourselves by recording the non-zero bilinears in the case of LLM i.e when $\e_- = - \g_5 \e_+$. For LLM only the following scalars
\be
\label{LLMscalars}
T_3 = i S_2 = i, \quad S_3 = -T_2 = \sinh \zeta,
\ee
and vectors
\be
\label{LLMvecs}
K^1 = K^8 = - \tfrac{1}{2} K, \quad K^3 = K^7 = -\tfrac{1}{2m} L, \quad K^6 = K^{10} = -\omega,
\ee
are non-zero. Here we have written the non-zero bilinears in the notation of \cite{LLM}.

We begin by using (\ref{const4}) and the symmetry properties of the gamma matrices to derive some constraints linking the various scalars\footnote{Here we employ the notation $A \cdot B \equiv A_{\mu} B^{\mu}$ and henceforth $\Re$ and $\Im$ will denote real and imaginary parts of expressions.}:
\bea
\label{alg1} 2 m S_2 + e^{-A} \Re(S_3) &=& 0, \\ \label{alg2}
\Im(S_3) &=& 0, \label{alg3} \\
\label{alg4} 4 m T_1 + e^{-6 \lambda -2 A} \mathcal{I} \Im(T_3) &=& 0, \\
\label{alg6} K^1 \cdot d (3 \lambda +A) &=& 0, \\
2 T_1 -  e^{-6 \lambda -A} \mathcal{I} T_2 &=& 0, \label{alg8} \\
\label{alg9} 2 e^{-A} T_2 - e^{-6\l -2 A} \mathcal{I} T_1 - 4 m \Im(T_3) &=& 0, \\
U_1 &=& 0.
\eea
As with all the relationships presented in this note, a valuable consistency check is provided by checking against LLM.

However, the important point not to be overlooked here is that without doing much work, we have already arrived at a different perspective of how the KSE constrains the geometry. If one follows the analysis of LLM, one is left with the impression that one first needs to set $\mathcal{I} = 0$ before one imposes $\e_- = - \g_5 \e_+$ to simplify the KSE. However,  by rewriting the information inherent in the algebraic constraints in terms of bilinears, we immediately see that $\mathcal{I} = 0$ \textit{follows from the relationship between the spinors.} This is evident from (\ref{alg8}) where $T_1$ is zero for LLM, while $T_2$ is non-zero.

In addition, the following list can be found from manipulating (\ref{const1}): \bea \Re(T_3)\mathcal{I}&=&0,\\
T_2\mathcal{I}&=&0,\\
K^1 \cdot d \lambda&=&0.
\eea

At this early stage, we are also seeing evidence of the direction $K^1$ emerging as a Killing direction, since it may be noted that the warp factors $\lambda$ and $A$ are independent of $K^1$. Combining some of the above equations, it is possible to infer that \be T_1=\mathcal{I}\Im(T_3)=0.\ee

So far, we have made use of the algebraic constraints, but one can also use the differential equation to write out torsion conditions for the scalars
\bea
\label{0tor1}
d( e^{-A} S_1) &=& -\tfrac{1}{2} e^{6 \l -3 A} \mathcal{I} K^3 + e^{-2 A} K^4, \\
d S_2 &=& -\tfrac{1}{2} e^{6 \l -2 A} \mathcal{I} K^4, \\
d(e^{-A} T_2) &=& 0, \\
\label{0tor4} d ( e^{-A} \Re(T_3)) &=& - e^{-2 A} \Im(K^7), \\
d \Im(T_3) &=& 0, \\
d U_2 &=& = - m K^5,
\eea
which are all satisfied for LLM. From here we see that $\Im(T_3)$ is a constant, which we henceforth denote $t$. Similarly, one can derive algebraic and differential relationships for the vectors. Despite only having four directions, we see from the text that one can define ten vectors. As they cannot all be independent, we note the following relations for the vectors:
\bea
\label{vecrel1} \mathcal{I}K^{10}&=&0, \\
\label{vecrel2} T_2d(3\lambda+A)+2m\Re(K^7)&=&0,\\
\label{vecrel4} -\Re(T_3)d(3\lambda+A)+e^{-A}\Im(K^7)+2mK^2&=&0,\\
-2S_1d(3\lambda+A)+e^{-6\lambda-2A}\mathcal{I}K^3-2e^{-A}K^4+4m\Im(K^8)&=&0,\\
\label{vecid5} 2S_2d(3\lambda+A)-e^{-6\lambda-2A}\mathcal{I}K^4+2e^{-A}K^3&=&0,\\
\label{vecrel7} 2\Im(S_3)d(3\lambda+A)+e^{-6\lambda-2A}\mathcal{I}\Re(K^8)&=&0,\\
-2\Re(S_3)d(3\lambda+A)+e^{-6\lambda-2A}\mathcal{I}\Im(K^8)+4mK^3&=&0, \\
2mK^5 - ie^{-A}K^9- U_2d(3\lambda+A) &=& 0.
\eea
One may also derive the torsion conditions for the vectors which we record in the appendix. These vector torsion conditions complete the geometric data required to construct a geometry by relating the fluxes $\mathcal{F}$ and $\mathcal{I}$ to vectors appearing in the metric.

\subsection{Killing directions}
The Killing directions from the possible list of vectors $K^{A}$ may be determined by using (\ref{diff3}) to determine $\nabla_{\mu} K_{\nu}^{A}$ and inspecting whether the symmetric part vanishes or not. As was observed in \cite{ads3M6}, when one treats the Killing spinors on an independent footing, it is not too surprising to find additional Killing directions. It may be confirmed that the two solutions to the Killing equation
\be
\nabla_{(\mu} K^{A}_{\nu)} = 0
\ee
are $K^1$ and $\Re(K^8)$. We stress that this is a general result, irrespective of whether $\mathcal{I}$ is zero or not, though $\Re(K^8)$ requires $\Im(S_3) = 0$ for it to be a solution. Fortunately, this is the case as may be confirmed from (\ref{alg3}).

We now show that both these directions are isometries of the full solution by showing that the Lie derivatives $\mathcal{L}_{X} \equiv i_{X} d + d i_X $ of the warp factors $\lambda$ and $A$ and the flux terms $\mathcal{F}$ and $\mathcal{G}$ with respect to $X \in \{K^1, \Re(K^8) \}$ are zero.  Firstly it is easy to show from any of (\ref{const1}), (\ref{const2}) or (\ref{const4}) that
\be
\label{scalarkill}
\mathcal{L}_{K^{1}} \lambda = \mathcal{L}_{K^{1}} A = 0,
\ee
and similarly for $\Re(K^8)$.

Now we would like to show that the Lie derivative of the flux term $\mathcal{F}$ is also zero w.r.t. $K^1$ and $\Re(K^8)$. It is easy to show using (\ref{const2}) that \bea
i_{K^{1}} \mathcal{F} &=& -2 e^{3 \lambda +2 A} \left[ \Re(S_3) d A + m K^3 \right], \nn
i_{\Re(K^{8})} \mathcal{F} &=& 2  e^{3 \lambda +A} \left[ T_2 d A - m \Re(K^7) \right].
\eea
Taking a derivative and using (\ref{alg1}), (\ref{vecrel2}) and (\ref{vecid5}) we find the following
\bea
\mathcal{L}_{K^1} \mathcal{F} &=& - 2m e^{-A} d (3 \lambda +A) \wedge K^{3}, \\
\mathcal{L}_{\Re(K^8)}  \mathcal{F} &=& 0.
\eea
Though the second line might be expected, the first line which is trivially zero for LLM has to be set to zero if we are to consider $K^1$ to be a symmetry of the complete solution. This in turn tells us that $d (3 \lambda +A)$ is a form in the $K^3$ direction. After eliminating the possibility of non-zero $\mathcal{I}$ in the next section, $K^1$ will become an isometry of the full solution. This may be seen clearly from (\ref{vecid5}) where with $\mathcal{I} =0$ and non-zero $S_2$, $d (3 \lambda +A)$ is along the $K^3$ direction. 

Finally, one may want to show that the Lie derivative of $\mathcal{G} = e^{-3 \lambda -2 A} \mathcal{I} vol(\mathcal{M}_4)$ is zero. This may be easily done by taking the Hodge dual, in which case, the result is immediate using the fact that $\mathcal{I}$ is constant and the warp factors are independent of the Killing directions (\ref{scalarkill}).

\section{No geometries with $\mathcal{I}$}
Having established some algebraic and differential relationships between the bilinears arising from the KSE, we are now in a fit position to comment on this case. When $\mathcal{I}$ is non-zero, we necessarily have to set the following bilinears to zero:
\bea
T_1 &=& \Im(S_3) = U_1 = 0 \label{genconst} \\
T_2 &=& T_3 = K^2 = K^7 = \Re(K^8) = K^{10} =  0 \label{Iconst1}.
\eea
The first line here is true in general i.e. for all geometries, while the second line follows as a result of insisting that $\mathcal{I}$ be non-zero. In deducing that the above vectors are zero, we have utilised (\ref{0tor4}), (\ref{vecrel1}), (\ref{vecrel2}), (\ref{vecrel4}) and (\ref{vecrel7}). In retaining $\mathcal{I}$, we have been forced to set to zero bilinears that are non-zero for LLM, so there is no way to reconcile the LLM class of solutions in \cite{LLM} with $\mathcal{I}$.

It is also worth noting that through naive counting, we have 16 real components of $\e_+, \e_-$ to be determined, while we have five scalar and four vector constraints making a total of twenty-one real constraints, therefore suggesting the system is overconstrained. Indeed, it is a straightforward, but lengthy exercise in linear algebra to show that the only solution to the above constraints is $\e_+ = \e_- = 0$. This may be most easily done by introducing an explicit representation for the gamma matrices as in the appendix and then writing $\e_+$ and $\e_-$ in terms of general complex components. We are thus led to the following conclusion:
\begin{quote}
\textit{There are no supersymmetric geometries within the LLM warped product ansatz with non-zero $\mathcal{I}$. }
\end{quote}

This unequivocal statement improves on the analysis in \cite{LLM} where it was shown that $\mathcal{I}$ could not be turned on perturbatively from the known class of solutions. This did not rule out the existence of $\mathcal{I}$ completely as we have done here.

\section{Geometries generalising LLM}
In this section we turn our focus to the relationship between the spinors $\e_- = - \g_5 \e_+$ to investigate to what extent it is determined by supersymmetry. One naively imagines, if one relaxes it, there is a small chance that more bilinears may be turned on in a way that leads to a generalisation. To extract sufficient information about the geometry to make a definite statement, we opt to initially study the algebraic Fierz identities.

Following \cite{M5} one can expand in terms of another basis of generic spinors $\eta_{a} \in \{ \e_{+}, \e_- \e_+^c, \e_-^c\}$ and construct the identity operator
\be
\mathbf{1} = \eta_{a} (m^{-1})^{ab} \bar{\eta}_b,
\ee
where $m_{ab} = \bar{\eta}_{a} \eta_{b}$. One may now use this completeness relation and (\ref{const4}) with $\mathcal{I} = 0$ to write
\be
\g^\mu\partial_\mu (3\lambda+A)= \Sigma_{a}^{~c} (m^{-1})^{ab} \eta_{c} \bar{\eta}_{b}.
\ee
Performing Fierz identities on $\eta_{c} \bar{\eta}_{b}$ one finds that the $\g_{\mu}$ coefficient gives the following useful relationship:
\bea
\label{diffAl}
d(3 \lambda +A) \left[S_2^2-S_1^2+|S_3|^2 \right] &=& \left\{e^{-A} S_1 K^4 + (2m S_3 - e^{-A} S_2) K^3 - 2 m S^1 \Im(K^8) \right\} \nn
\eea
Note the $1$ and $\g_5 \g_{\mu}$ coefficients recover (\ref{alg9}) and (\ref{alg1}) respectively, while $\g_5$ coefficient is trivially satisfied. One may also extract information from the $ \g_{\mu \nu}$ coefficient, but the relationship is in terms of 2-forms which we find less useful. Moreover, one may repeat with the other algebraic constraints from the KSE, but one recovers known relationships for the scalars and vectors.

Note also now for $K^1$ to be an isometry of the overall solution, we require $d(3 \l +A) \wedge K^3 =0$ which as we have seen is true when $\mathcal{I} = 0$. This in turn implies from (\ref{diffAl}) that
\be
\label{lincomb}
s K^4 - 2 m e^{A} \Im(K^8) = S_1 K^3,
\ee
where we have determined the factor on the right hand side by using the vector identity (\ref{vecrel4}). The outstanding independent vector identities may then be expressed as
\bea
\label{othervecs}
s d (3 \lambda +A) + e^{-A} K^3 &=& 0, \nn
s \Re(K^7) - t K^3 &=& 0, \nn
\Re(T_3) K^3 + s \Im(K^7) + 2 m s e^{A} K^2 &=& 0, \nn
U_2 K^3 + 2 m s e^{A} K^5 - i s K^9 &=& 0,
\eea
where we have eliminated factors of $ d(3 \lambda +A)$ which are non-zero in general. As $S_2$ is a constant when $\mathcal{I} = 0$, in addition to the constant $\Im(T_3) = t$, we also have ${S_2} = s$. Both these constants we take to be real and non-zero as is the case in LLM.

With the above relations between the vectors (\ref{lincomb}) and (\ref{othervecs}) derived from supersymmetry, we are now in an adequate position to address whether there is a generalisation within the LLM class of solutions. Some details may be found in appendix B, so here we just present an overview of the arguments.

We now consider $\e_+$ to be generic as in LLM, but allow for it to be related to $\e_-$ through complex functions $a, b, c, d$
\be
\e_- = a \e_+ + b \g_5 \e_+ + c \e_+^{c} + d \g_5 \e_+^c.
\ee

Note as we are only expecting one Killing direction from $\mathcal{M}_4$ corresponding to the $R$-symmetry\footnote{Typically in $AdS \times M$ geometries with enough supersymmetry, if an $R$-symmetry is expected, then
it must come out from the Killing spinors.}, it is most natural that $K^1$ and $\Re(K^8)$ should be the same\footnote{They can be related up to a constant and the result is unchanged.}. As we have seen earlier in (\ref{LLMvecs}), this is true for LLM. Note also from (\ref{LLMvecs}) that setting one of these vectors to zero, takes one out of the LLM class of solutions. In addition, one may imagine that some linear combination of these two vectors could be an $R$-symmetry direction, while another may correspond to a normal isometry direction. It would be interesting to explore this option, but it would take us out of the LLM class, so we do not consider this possibility here.

The strategy then is to simply plug this expression for $\e_-$ into the above vector conditions, while also taking into account the scalar constraints:
\bea
\label{sec4scalar}
T_1 &=& \Im(S_3) = U_1 = 0 , \nn
 0 &=& 2 m s+ e^{-A} S_3, \nn
 0 &=& T_2 - 2m e^{A} t.
\eea

One then finds a series of equations in terms of the original scalars and vectors ($K, L, \omega$) of LLM - we are assuming these are not trivially zero, so that any solution to the KSE will include the solution presented in \cite{LLM}. As it is possible to use the Fierz identities to show that $K, L$ and $\omega$ are orthogonal, $K \cdot L = K \cdot \omega = L \cdot \omega = 0$, one can use them as a basis to expand the vector identities above. By identifying their coefficients, one is lead to a collection of algebraic equations on $a, b, c$ and $d$.

It is possible to quickly see that the condition $K^1 = \Re(K^8)$ is very strong. In terms of component vectors one finds the following relationships among $a,b,c,d$:
\bea
\label{identify}
\Re(b) &=& -\frac{1}{2} ( |a|^2+|b|^2+|c|^2+|d|^2+1), \nn
\Re(a) &=& - \Re(a^*b-c^*d), \nn
d &=& -(a^*c+b^*d).
\eea
Then using the vector torsion identities (\ref{vec1}) and (\ref{vec8}) and the fact that terms proportional to $m$ and $\mathcal{F}$ will in general vanish independently, one finds another set of identities:
\bea
s &=& t, \nn
\Re(b) &=& -\frac{1}{2} ( 1 - |a|^2+|b|^2-|c|^2+|d|^2), \nn
\Im(a) &=& \Im(a^*b + c^* d), \nn
c &=& a^* d - b^* c, \nn
d &=& a^* c - b^*d.
\eea
We can reconcile these two only when $a = c = 0$. Now, it is a simple task to use the scalar constraints as rewritten in appendix B to see that the only solution is
\be
a = c= d= 0, \quad  b = -1.
\ee

In other words, it appears that the most general relationship between the Killing spinors permitted by supersymmetry is $\e_- = - \g_5 \e_+$, so it appears that there is no way to generalise the LLM solution when $\mathcal{I} = 0$ by relaxing the original relationship assumed in \cite{LLM}. Interestingly, this result is an immediate consequence of identifying the two Killing directions $K^1$ and $\Re(K^8)$. It is certainly interesting to see if there are any geometries with non-parallel $K^1$ and $\Re(K^8)$. It is possible that such supersymmetric geometries are all maximally supersymmetric.

In conclusion, we have examined the assumptions leading to the LLM class of solutions in M-theory using Killing spinor bilinears and the KSE. In this language it is easy to see that there are no supersymmetric solutions with $\mathcal{I}$ and when one sets this flux term to zero, there is no more general class of LLM solutions.

\section*{Acknowledgements}
We are grateful to Oleg Lunin for helpful and supportive comments both at the start and end of this project. We would like to acknowledge discussion on a tangent direction with Dario Martelli and Bogdan Stefanski, while we thank Jerome Gauntlett, Ki-Myeong Lee and Piljin Yi for sharing their time and experience. JW would like to thank the Center for High Energy Physics, Peking University for hospitality and Bin Chen for discussions on a recent visit.

\appendix

\section{Conventions}
We will be borrowing our conventions from \cite{LLM} where we refer the reader for details. Here we provide a brief summary. On the external space $\mathcal{M}_4$ with signature $(-,+,+,+)$ we will take $\e^{0123} = 1$. As a result, defining  $\g_5=i\g_{0123}$, we have $ \g_{5}^2 = +1$. We will then adopt
\bea
\label{pc}
(\g_0)^\dagger=-\g_0, \quad (\g_i)^\dagger=\g_i, \eea
Also from  LLM \cite{LLM}, we see that the intertwiners $A$ and $C$ are given by
\be
A \equiv \g^{0}, \quad C \equiv \g^2.
\ee
From (\ref{pc}) this means that
\be
A \g_{\mu} A^{-1} = - \g_{\mu}^{\dagger}. \quad
\ee
In LLM $\g^2$ is antisymmetric and $\g^{0}, \g^{1}, \g^{3}$ symmetric so that
\be
C^{-1} \g_{\mu}^{T} C = - \g_{\mu}.
\ee
Note that subject to these choices
\be
\g_5^{\dagger} = \g_5, \quad \g_5^{T} = - \g_5.
\ee

Then defining $D = C A^{T}$ in the usual fashion, we can define the conjugate spinor to $\e$ as $\e^{c} = D \e^{*} = \g^{2} \g^{0} \e^{*}$. This implies that $\bar{\e}^{c} = - \e^{T} \g^{2}$. Note also that $D= \g^{2} \g^{0}$ and that $D D^* = +1$, so that we have the freedom to take $\e$ to be a Majorana spinor provided we impose $\e^{c} = \e$.

Given spinors $\chi, \xi$ and spinor bilinears constructed from $p$ antisymmetrised gamma matrices $\g^{(p)} \equiv \g^{\mu_1 \cdots \mu_p}$, we have the following symmetry properties for the spinor bilinears
\bea
(\bar{\chi} \g^{(p)} \xi)^{\dagger} &=& (-1)^{\tfrac{p(p+1)}{2} +1} \bar{\xi}  \g^{(p)} \chi,  \nn
(\bar{\chi}^c \g^{(p)} \xi)^{T} &=& (-1)^{\tfrac{p(p+1)}{2} +1} \bar{\xi}^c  \g^{(p)} \chi.
\eea

Finally, an explicit representation of the above $\gamma$ matrices may be written:
\bea
\g_0 &=& 1 \otimes i \sigma_3, \quad
\g_1 = 1 \otimes \sigma_1, \quad
\g_2 = \sigma_1 \otimes \sigma_2, \quad
\g_3 = \sigma_2 \otimes \sigma_2.
\eea
Employing this choice $\g_5 \equiv i \g_{0123} = \sigma_3 \otimes \sigma_2$.


\section{Relaxing LLM Killing spinor relation}
In this appendix we present some relationships that are useful in determining $\e_- = - \g_5 \e_+$ from the identities we have derived in section 4, namely (\ref{lincomb}), (\ref{othervecs}) and (\ref{sec4scalar}). As explained in the text, given two generic spinors in four-dimensions, they can be related via
\be
\label{expand}
\e_-= a\e_+ + b \g_5 \e_+ + c\e_+^c + d \g_5 \e_+^c,
\ee
where $a, b, c, d$ are complex-valued functions.
Slotting this expression for $\e_-$ into the scalar equations (\ref{sec4scalar}), they may be rewritten as
\bea
\label{Beq1} && f_1 (-1+|a|^2-|b|^2-|c|^2+|d|^2)+2f_2\Im(a^*b-c^*d) -2s =0, \\
\label{Beq2} && if_1 c+ f_2 d=0, \\
&& iaf_1 +bf_2 +2m s e^A=0, \\
&& f_2-2mt e^A=0, \\
\label{Beq5} && f_2 \Im(a) + f_1 \Re(b) -t=0, \\
\label{Beq6} && f_2 (1+|a|^2-|b|^2-|c|^2+|d|^2)-2f_1\Im(a^*b-c^*d) =0.
\eea
Here we have introduced the added notation $\bar{\e}_+\e_+= -\bar{\e}_+^c\e_+^c = if_1$ and $ \bar{\e}_+\g_5\e_+= -\bar{\e}_+^c\g_5\e_+^c= f_2$, with $f_i \in \mathbb{R}$.  Note that (\ref{Beq2}) to (\ref{Beq5}) may be satisfied if
\be \label{b} b=-\frac{s}{t}+i\tilde{d}a,  \ee
where we have defined a new real function $\tilde{d} = - f_1/f_2$. Note that this function is non-zero in LLM.

Moving along to the vector expressions, (\ref{lincomb}) and (\ref{othervecs}), we can write all the vector expressions in terms of the vectors in LLM \cite{LLM} as they constitute a basis. They may be written
\bea
&& K_{\mu} =-2\bar{\e}_+ \g_{\mu} \e_+ = -2\bar{\e}_+^c \g_{\mu} \e_+^c \\
&& L_{\mu} =-2m \bar{\e}_+ \g_5 \g_{\mu} \e_+ = 2m \bar{\e}_+^c \g_5 \g_{\mu} \e_+^c \\
&& \omega= -\bar{\e}_+^c \g_{\mu} \e_+
\eea
It is also easy to show that $\omega^* = - \bar{\e}_+ \g_{\mu} \e_+^c$ and that
\be
K \cdot L=K \cdot \omega = L \cdot\omega =0,
\ee
so they are all orthogonal and constitute a basis. It is useful to document all the vectors in our text $K^A$  in terms of  the original LLM vectors and $a,b,c,d$ as follow
\bea
&& K_1+K_2= -\frac{1}{2}K\\
&& K_1-K_2 = -\frac{1}{2}(|a|^2+|b|^2+|c|^2+|d|^2)K + \frac{1}{m}\Re(a^*b-c^*d) L \\
&& \hspace{20mm} - 2 \Re\left((a^*c+b^*d) \omega^* \right) \\
&& K_3+K_4 =-\frac{1}{2m}L \\
&& K_3-K_4 =\Re(a^*b+c^*d)K -\frac{1}{2m} (|a|^2+|b|^2-|c|^2-|d|^2)L\\
&& \hspace{20mm} + 2\Re\left((a^*d+b^*c)\omega^* \right) \\
&& K_5+K_6 =-\omega \\
&& K_5-K_6 =(bd-ac)K +\frac{1}{m}(bc-ad)L+(b^2-a^2)\omega +(d^2-c^2) \omega^*  \\
&& K_7 = -\frac{1}{2}aK+\frac{1}{2m} bL-c\omega^* \\
&& K_8 =\frac{1}{2} bK - \frac{1}{2m}aL+d\omega^* \\
&& K_9=-\frac{1}{2}cK-\frac{1}{2m}dL -a\omega\\
&& K_{10} =\frac{1}{2}dK+\frac{1}{2m}cL+b\omega.
\eea

\section{Vector torsion conditions}
In this section we illustrate the result of converting the KSE equations into differential conditions on the various vectors. Some of the expressions below allow one to determine an expression for the two-form flux term $\mathcal{F}$, thus completing the information about the geometry. By using the KSE and the algebraic constraints on occasion, one can rewrite the following vector torsion conditions:
\bea
\label{vec1} d K^{1} &=& e^{-6 \lambda -2 A} \mathcal{I} \tfrac{i}{4} \left[\bar{\e}_{+} \g_5 \g_{\mu  \nu} \e_+ +  \bar{\e}_{-} \g_5 \g_{\mu  \nu} \e_- \right]  dx^{\mu \nu} \nn &+& \tfrac{m}{2} \left[ \bar{\e}_{+} \g_5 \g_{\mu  \nu} \e_- -  \bar{\e}_{-} \g_5 \g_{\mu  \nu} \e_+ \right] dx^{\mu \nu} - e^{-3 \lambda -2 A} \Re(S_3) \mathcal{F}, \\
\label{vec2} d ( e^{3 \lambda} K^2) &=& e^{-3 \lambda -2 A} \mathcal{I} \tfrac{i}{8} \left[ \bar{\e}_{+} \g_5 \g_{\mu  \nu} \e_+ -  \bar{\e}_{-} \g_5 \g_{\mu  \nu} \e_- \right] dx^{\mu \nu}, \\
\label{vec3} d(e^{-A} K^3) &=& 0, \\
\label{vec4} d K^4 &=& \tfrac{m}{2} \left[\bar{\e}_{+} \g_{\mu \nu} \e_{-} + \bar{\e}_{-} \g_{\mu \nu} \e_{+}\right] dx^{\mu \nu} - e^{-3 \lambda -2 A} \Re(T_3) \mathcal{F}, \\
\label{vec5} d K^5 &=& e^{-6 \lambda -2 A} \mathcal{I} \tfrac{i}{4} \left[ \bar{\e}_{+}^c \g_5 \g_{\mu  \nu} \e_+ +  \bar{\e}_{-}^c \g_5 \g_{\mu  \nu} \e_- \right] dx^{\mu \nu}, \\
\label{vec6} d (e^{3 \lambda} K^{6}) &=& e^{-3 \lambda -2 A} \mathcal{I} \tfrac{i}{8} \left[ \bar{\e}_{+}^c \g_5 \g_{\mu \nu} \e_{+} - \bar{\e}_{-}^c \g_5 \g_{\mu \nu} \e_{-}\right] dx^{\mu \nu}  \nn
&-& e^{3 \lambda} \tfrac{m}{4} \left[\bar{\e}_{+}^c \g_5 \g_{\mu \nu} \e_{-} + \bar{\e}_{-}^c \g_5 \g_{\mu \nu} \e_{+} \right] dx^{\mu \nu}, \\
\label{vec7} d K^7 &=& e^{-6 \lambda -2 A} \mathcal{I} \left[ \tfrac{i}{2} \Re(\bar{\e}_{+} \g_5 \g_{\mu \nu} \e_{-}) -  \tfrac{1}{4} \Im(\bar{\e}_{+} \g_5 \g_{\mu \nu} \e_{-})\right] dx^{\mu \nu}\nn
&-& \tfrac{m}{2} \left[ \bar{\e}_{+} \g_5 \g_{\mu  \nu} \e_+ +  \bar{\e}_{-} \g_5 \g_{\mu  \nu} \e_-\right] dx^{\mu \nu}  - 3 d \lambda \wedge \Re(K^7) \nn &-& i S_1 \mathcal{F}, \\
\label{vec8} d K^8 &=& -\tfrac{m}{2} \left[ \bar{\e}_{+} \g_{\mu  \nu} \e_+ -  \bar{\e}_{-} \g_{\mu  \nu} \e_-\right] dx^{\mu \nu}  - 3 i d \lambda \wedge \Im(K^8) \nn &+& e^{-3 \lambda -2 A} T_2 \mathcal{F} , \\
\label{vec9} d(e^{3 \l} K^9) &=& e^{-3 \lambda -2 A} \mathcal{I} \tfrac{i}{4} \bar{\e}^c_{+} \g_5 \g_{\mu \nu} \e_- dx^{\mu \nu}  \nn &+& e^{3 \lambda} \tfrac{m}{4} \left[  \bar{\e}_{+}^c \g_5 \g_{\mu  \nu} \e_+ -  \bar{\e}_{-}^c \g_5 \g_{\mu  \nu} \e_- \right] dx^{\mu \nu}, \\
\label{vec10}  d(e^{-A} K^{10}) &=& - e^{-A} \tfrac{m}{4} \left[  \bar{\e}_{+}^c  \g_{\mu  \nu} \e_+ +  \bar{\e}_{-}^c \g_{\mu  \nu} \e_-\right] dx^{\mu \nu}  + e^{-2 A} \tfrac{i}{2} \bar{\e}^{c}_{+} \g_{\mu \nu} \e_- dx^{\mu \nu}.
\eea

Note (\ref{vec1}) \& (\ref{vec8}), (\ref{vec3}) \& (\ref{vec7}), (\ref{vec6}) are consistent with (F.25), (F.30), (F.70) of LLM \cite{LLM} respectively. In contrast, (\ref{vec2}), (\ref{vec4}), (\ref{vec5}), (\ref{vec7}), (\ref{vec9}) are trivial for LLM. There is a factor of two difference between our $\mathcal{F}$ and that of LLM. Also, we have made use of the following form notation:
\bea
A &\equiv& \tfrac{1}{p!} A_{i_1 \cdots i_p} dx^{i_1 \cdots i_p,}~~~\mbox{($p$-form)}, \nn
dx^{i_1 \cdots i_p} &\equiv& dx^{i_1} \wedge dx^{i_2} \wedge \cdots \wedge dx^{i_p}.
\eea


\begin{thebibliography}{99}
\bibitem{LLM}
  H.~Lin, O.~Lunin and J.~M.~Maldacena,
  ``Bubbling AdS space and 1/2 BPS geometries,''
  JHEP {\bf 0410} (2004) 025
  [arXiv:hep-th/0409174].

\bibitem{GM}
  D.~Gaiotto and J.~Maldacena,
  ``The gravity duals of N=2 superconformal field theories,''
  arXiv:0904.4466 [hep-th].

\bibitem{M6}
  J.~P.~Gauntlett, D.~Martelli, J.~Sparks and D.~Waldram,
  Class.\ Quant.\ Grav.\  {\bf 21}, 4335 (2004)
  [arXiv:hep-th/0402153].

  \bibitem{M5}
  J.~P.~Gauntlett, D.~Martelli, J.~Sparks and D.~Waldram,
  ``Supersymmetric AdS(5) solutions of type IIB supergravity,''
  Class.\ Quant.\ Grav.\  {\bf 23}, 4693 (2006)
  [arXiv:hep-th/0510125].

\bibitem{ads3M6}
  E.~\'O~Colg\'ain, J.~B.~Wu and H.~Yavartanoo,
  ``Supersymmetric AdS3 X S2 M-theory geometries with fluxes,''
  arXiv:1005.4527 [hep-th].

\bibitem{Donos}
  A.~Donos and J.~Simon,
  ``The electrostatic view on M-theory LLM geometries,''
  arXiv:1010.3101 [hep-th].

\bibitem{int}
  J.~P.~Gauntlett and S.~Pakis,
  ``The geometry of D = 11 Killing spinors,''
  JHEP {\bf 0304}, 039 (2003)
  [arXiv:hep-th/0212008].

\end{thebibliography}
\end{document}